\documentclass[superscriptaddress, twocolumn,amsmath,amssymb,prl]{revtex4-1}
\usepackage{graphicx}
\usepackage{dcolumn}
\usepackage{multirow}
\usepackage{booktabs}
\usepackage{bm,color}
\usepackage{braket}

\begin{document}

\title{Photogalvanic Effect in Weyl Semimetals from First Principles}

\author{Yang Zhang}
\affiliation{Max Planck Institute for Chemical Physics of Solids, 01187 Dresden, Germany}
\affiliation{Leibniz Institute for Solid State and Materials Research, 01069 Dresden, Germany}

\author{Hiroaki Ishizuka}
\affiliation{Department of Applied Physics and Quantum Phase Electronics Center (QPEC), University of Tokyo, Tokyo 113-8656, Japan}

\author{Jeroen van den Brink}
\affiliation{Leibniz Institute for Solid State and Materials Research, 01069 Dresden, Germany}

\author{Claudia Felser}
\affiliation{Max Planck Institute for Chemical Physics of Solids, 01187 Dresden, Germany}

\author{Binghai Yan}
\affiliation{Department of Condensed Matter Physics, Weizmann Institute of Science, Rehovot 7610001, Israel}

\author{Naoto Nagaosa}
\affiliation{RIKEN Center for Emergent Matter Science (CEMS), Wako, 351-0198, Japan}
\affiliation{Department of Applied Physics and Quantum Phase Electronics Center (QPEC), University of Tokyo, Tokyo 113-8656, Japan}

\begin{abstract}
Using first-principles calculations, we investigate the photogalvanic effect in the Weyl semimetal material TaAs.
We find colossal photocurrents caused by the Weyl points in the band structure in a wide range of laser frequency.
 Our calculations reveal that the photocurrent is predominantly contributed by the three-band transition
 from the occupied Weyl band to the empty Weyl band via an intermediate band away from the Weyl cone, for excitations
 both by linearly and circularly polarized lights.
 Therefore, it is essential to sum over all three-band transitions by considering a full set of Bloch bands
 (both Weyl bands and trivial bands) in the first-principles band structure while it does not suffice to only
 consider the two-band direct transition within a Weyl cone.
The calculated photoconductivities are well consistent with recent experiment measurements.
 Our work provides the first first-principles calculation on nonlinear optical phenomena of Weyl semimetals and serves
 as a deep understanding of the photogalvanic effects in complexed materials.
\end{abstract}

\maketitle

\textit{Introduction.}
  Weyl fermions correspond to the massless solutions of Dirac equation~\cite{weyl1929elektron} and have been observed
  in solids as quasiparticles recently~\cite{Weng2015,Huang2015,Lv2015TaAs,Xu2015TaAs,Yang2015TaAs}.
  Related materials are called Weyl semimetals
  (WSM)~\cite{Wan2011,volovik2003universe,murakami2007phase,Burkov2011,Hosur2013,Yan2017,Armitage2017}.
 A WSM gives rise to linearly band-crossing points called Weyl points (WPs) in the momentum space.
 WPs are monopoles of the Berry curvature~\cite{Nagaosa2010,Xiao2010} with finite chirality and are related to the chiral anomaly in the context of high-energy physics~\cite{Nielsen1981,Nielsen1983,Xiong2015,Gooth2017} and unique surface Fermi arcs~\cite{Weng2015}.

 The monopole-type Berry curvature of WSMs can lead to appealing nonlinear optical effects that are intimately related to the Berry phase in the
 band structure
 \cite{Moore2010,Deyo2009,young2012first,Sipe2000,morimoto2016topological}.
 Under strong light irradiation, an noncentrosymmetric material exhibits
 photocurrents as nonlinear functions of the electric field of the light and also generates higher harmonic frequencies,
 referred to as the photogalvanic effects.
 The photogalvanic effect rectifies light to dc currents and often play a crucial role
 in optical devices and solar cells beyond the p-n junction platform~\cite{Choi2009,Yang2010,Grinberg2013}. Under linearly polarized light, the induced photocurrent is usually called shift current that originates
 in the charge center shift between the valence and conduction bands in the optical excitation. Under the circularly polarized light, the photocurrent generation
 is referred to as the circular photogalavnic effect (CPGE).
 It can be expressed in the formalism of Berry curvature and Berry
 connection~\cite{young2012first,Sipe2000,morimoto2016topological}, revealing a topological nature.
 Therefore, WSMs have recently been theoretically investigated for such nonlinear optical
 phenomena~\cite{Vazifeh2013,Goswami2015,Kargarian2015,ishizuka2016emergent,ishizuka2017momentum,Hosur:2011gf,sodemann2015quantum,
 chan2016chiral,morimoto2016semiclassical,Taguchi:2016ef,de2017quantized,chan2017photocurrents,
 Konig:2017gy,Rostami:2017ul,Golub:2017dk,zhang2018berry,Yang:2017up}.  In these works, two-band or four-band
  effective models are commonly adopted to reveal the relation between the
 photocurrent and the Weyl bands. For example, the tilt of Weyl cones is proposed to play an
 essential role to generate a net CPGE current by considering the two-band transition from the occupied Weyl band to the empty Weyl band~\cite{chan2017photocurrents}.
 However, the first-principles investigation on the photogalvanic effects of WSMs, which accounts for the realistic material band structures, is still missing.

Recent experiments~\cite{wu2017giant,ma2017direct,sun2017circular,osterhoudt2017colossal,lim2018temperature,ji2018spatially} have reported giant photocurrents effects and the second-harmonic generation (SHG) in the TaAs-family WSMs exhibiting in orders of magnitude larger responses than conventional nonlinear materials.
 However,  some experiments are seemingly controversial to each other. Reference~\onlinecite{ma2017direct} reported a photocurrent caused by the circularly polarized light, but claimed that a negligible photocurrent was caused by the linearly polarized light through the shift current mechanism. In contrast, Ref.~\onlinecite{osterhoudt2017colossal} reported a colossal shift current with linearly polarized light in the same compound. Therefore, accurate estimations of photocurrents are necessary and timely to identify quantitative contributions from CPGE and shift current for a specific material.
 In addition, nonlinear optical phenomena are highly sensitive to the bulk Fermi surface topology but are insensitive to surface states. Hence, they can serve a direct pathway to probe the topology inside the bulk.

\begin{figure}[htbp]
\begin{center}
\includegraphics[width=0.35\textwidth]{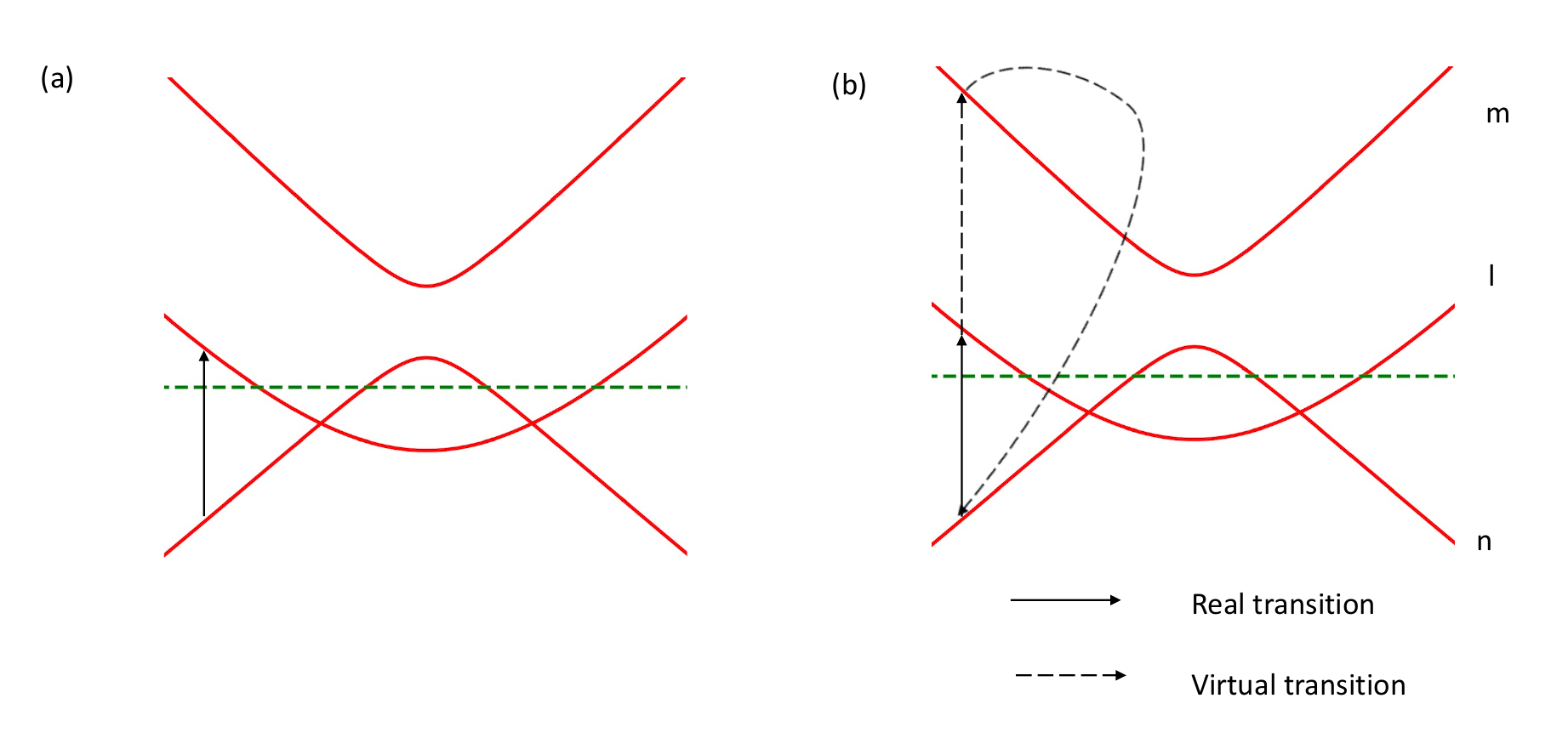}
\end{center}
\caption{
Optical process for bands with a pair of Weyl nodes of
(a) absorption, (b) dc photocurrent.
}
\label{optical}
\end{figure}

  In this letter, we perform first-principles studies on the CPGE and shift current effect in WSMs.
  With the second-order Kubo formulism, we calculate the photocurrent conductivity in the inversion-asymmetric WSM
  TaAs via a multiband approach. Our results agree quantitatively with recent experiments.
  The shift current displays a close relation with the existence of WPs. Specially in the long-wavelength region,
  the shift current is predominantly contributed by virtual transitions from the occupied Weyl to the empty Weyl band
  through a third trivial band, referred to as the three-band transition, as illustrated in Fig.~\ref{optical}b.
 For CPGE, the three-band virtual transitions make the dominant contributions and distribute relatively uniformed in the momentum space.
 In contrast, the two-band real transitions contribute much less photocurrent, which is mainly caused by the Weyl cone regions.
Given the significance of the three-band transitions, it is necessary to sum over all intermediate states by consdiering a full set of Bloch states.
 Then the first-principles method is naturally the best way to compute the nonlinear response.
 For the same photon energy used in experiment, we find that the CPGE photocurrent is nearly two orders of magnitude greater than the shift current
 and clarify the possible reason why the shift current was not detected in a previous experiment that reported the CPGE~\cite{ma2017direct} .

\textit{Theory and method.}
 The calculation of CPGE and shift current is based on a quadratic response
 theory proposed by von Baltz and Kraut~\cite{kraut1979anomalous,von1981theory,kristoffel1980some},
which accounts for the steady-state short-circuit photocurrent under the linearly polarized
 light. To calculate the photocurrent also for the circularly polarized light, we
 have generalized this quadratic response theory to a more general relation for
 the photoconductivity:
\begin{equation}
\label{second-kubo}
\begin{aligned}
  & \sigma^c_{ab}= \frac{|e|^3}{8\pi^3 \omega^2} Re \bigg\{ \phi_{ab}
  \sum_{\Omega=\pm \omega}
  \sum_{l,m,n} \int_{BZ} d^3k (f_l- f_n) \\
  & \frac{<n \vec{k}|\hat{v}_a|l \vec{k}><l
  \vec{k}|\hat{v}_b|m\vec{k}><m\vec{k}|\hat{v}_c|n\vec{k}>}
  {(E_n-E_m-i\delta)(E_n-E_l + \hbar \Omega- i\delta)} \bigg\}.
\end{aligned}
\end{equation}
  The conductivity ($\sigma^c_{ab};a,b,c=x,y,z$) is a third rank tensor and
  represents the photocurrent $J^c$ generated by an electrical field $\vec{E}$ via
  $J^c=\sigma^c_{ab}E_a^{*}E_b$.
  Here $\hat{v}_a=\frac{ \hat{p}}{m_0}$, $E_n=E_n(\vec{k})$,
  and $m_0$, $\delta=\hbar/\tau$, $\tau$
   stand for, respectively, free electron mass, broadening parameter, and the quasiparticle lifetime.
  $\phi_{ab}$ is the phase difference
  between driving field $\vec{E}_a$ and $\vec{E}_b$, i.e. $\phi_{yz}=i$ for left-circular
  polarized light propagating in $x$ direction with light-polarization vector
  $(0,1,i)$.
  It is clear that the real part of the integral in
  Eq.~\ref{second-kubo} describes the shift
  current response under linearly polarized light and the imaginary part of the integral
  gives the helicity dependent CPGE.

  Next, we analyze the response tensor under time reversal symmetry ($\hat{T}$) and the point group symmetry. For simple, we define
  $N \equiv <n \vec{k}|\hat{v}_a|l \vec{k}><l
  \vec{k}|\hat{v}_b|m\vec{k}><m\vec{k}|\hat{v}_c|n\vec{k}> $.
  $\hat{T}$ reverses the velocity and brings an additional minus sign to the imaginary part of $N
  $ by the complex conjugation. Thus, in materials with time reversal symmetry, the real
  part of the numerator is odd to $\vec{k}$ and therefore vanishes in the integral, and hence,
  only the imaginary part of the numerator has to be taken into
  account for calculations on non-magnetic WSMs. Since there is no current from $l=n$ or $m=n$,
  we can separate the contribution into two parts with respect to band number $l$ and $m$. The
  three-band processes ($n \rightarrow m \rightarrow l$) are given by $l \ne m$, and the two-band processes are given by
  $l=m$ (two-band transition). By applying the point group symmetry operations to
  the numerator $N$, the third rank conductivity tensor shape can
  be determined, as can be the tensor form of the anomalous Hall conductivity  and spin Hall conductivity ~\cite{vzelezny2017spin,zhang2017strong}.

  To see the relations between photocurrent response and the detailed band structure,
  we analyze the energy denominator by decomposing it into real and
  imaginary parts:
\begin{small}
\begin{equation}
\label{denominators}
\begin{aligned}
  & D_1=\frac{1}{E_n-E_m-i\delta}=\frac{P}{E_n-E_m}+i\pi\delta(E_n-E_m),\\
  & D_2=\frac{1}{E_n-E_l+\hbar\Omega-i\delta}=\frac{P}{E_n-E_l+\hbar\Omega}+i\pi\delta(E_n-E_l+\hbar\Omega), \\
\end{aligned}
\end{equation}
\end{small}
  Since the product of the three velocity matrices is purely imaginary,
  $Im(D_1D_2)$ ($\sim \pi\frac{P}{(E_n-E_m)}\delta(E_n-E_l+\hbar\Omega)$)
  gives the shift current response when $\phi_{ab}$ is real.
  Only the momentum vector with band gap equal to photon energy
  ($|E_n-E_l|=\hbar w$) contributes to the response under the linearly
  polarized light.
  It indicates that the shift current distribute mainly in some selective small areas in the momentum space.
   When the incident photon energy is sufficiently small, the
  response current only comes from the gap between two Weyl bands due to
  the energy selection rule. In the $\delta = \hbar / \tau \to0$ limit (long
  relaxation time limit, which is valid for semiconductors and insulators), the summation
  over band $m$ can be performed analytically via the first-order perturbation
  correction of Bloch-wave function~\cite{von1981theory}. In the end, we
  obtain the shift vector formula for the shift current density
  ~\cite{von1981theory,young2012first}. The shift vector directly
  connects the response photocurrent with a charge center shift between
  valance and conduction bands, but is quite numerically unstable for metallic system
  with a low-frequency driving field, due to the energy delta function and gauge fix of
  Berry connection of valence and conduction bands, and is not suited to deal
  with scattering processes with finite relaxation time.
  In a two-band approximation, the shift current response $\sigma^a_{aa}(a=x,y,z)$ is
  zero as the velocity numerator $N$ is real (here $l=m$, $N=<l\vec{k}|\hat{v}_a|l\vec{k}>|<n \vec{k}|\hat{v}_a|l \vec{k}>|^2$),
  in which the velocity $v_a \equiv <l\vec{k}|\hat{v}_a|l\vec{k}>$ is odd to $\vec{k}$ due to the time reversal symmetry.
 Therefore,  to calculate the shift current in real materials properly, one needs to use a multiband approach beyond the two-band approximation.

  For circularly polarized light with helicity dependent term $\phi_{ab}=i$, the dispersive part
  $Re(D_1D_2)$ ($\sim \frac{1}{(E_n-E_m)(E_n-E_l+\hbar\Omega)}$
  (note relaxation time plays a minor role in CPGE)
  contributes to the response photocurrent.
  The absence of $\delta$--function in $Re(D_1D_2)$ indicates that there is no specific energy selection rule in the transition.
   Thus, in contrast to the concentrated distribution of the shift current,
   the CPGE distribution can be rather smeared out in momentum space.
   It also indicates that different transition pathways (real and virtual) contribute relatively equally to the photocurrent, assuming comparable numerators $N$.
   Given the large number of three-band virtual transitions, the virtual process might overwhelm the two-band direct process to induce the photocurrent.

To calculate the second-order photoconductivity in realistic compounds, we
obtain the density-functional theory (DFT) Bloch wave functions from the Full-Potential Local-Orbital
program (\textsc{FPLO})~\cite{koepernik1999full} within the generalized gradient approximation (GGA)~\cite{perdew1996}.
By projecting the Bloch wave functions onto Wannier
functions, we obtain a tight-binding Hamiltonian with 32 bands, which we use for efficient
evaluation of the photocurrent.
For the integrals of Eq. \ref{second-kubo}, the BZ was sampled by $k$-grids
from $200\times200\times200$ to $960\times960\times960$. Satisfactory convergence
was achieved for a $k$-grid of size $240\times240\times240$ for all three compounds.
Increasing the grid  size to $960\times960\times960$ varied the
conductivity by less than 5\%.

\textit{Realistic materials.}
The material TaAs belongs to point group $4mm$, and has mirror reflections in $x$ and $y$ directions.
Due to the mirror symmetries, the nonzero conductivity elements are limited to the ones with an even number of $x$ and $y$, i.e.
$\sigma^z_{xx}, \sigma^z_{yy}, \sigma^z_{zz}, \sigma^y_{zy}(\sigma^y_{yz}),
\sigma^x_{zx}(\sigma^x_{xz})$.
In addition, the $4_2$ screw rotation symmetry about $z$ axis gives the relation
$\sigma^z_{xx}=\sigma^z_{yy}, \sigma^y_{zy}=\sigma^x_{zx}$. Therefore, only three
independent elements exist, i.e. $\sigma^z_{xx}$, $\sigma^y_{zy}$ and $\sigma^z_{zz}$. For the shift current,
all three elements matter. For CPGE, only $\sigma^y_{zy}$ is relevant.

\begin{figure}[htbp]
\begin{center}
\includegraphics[width=0.4\textwidth]{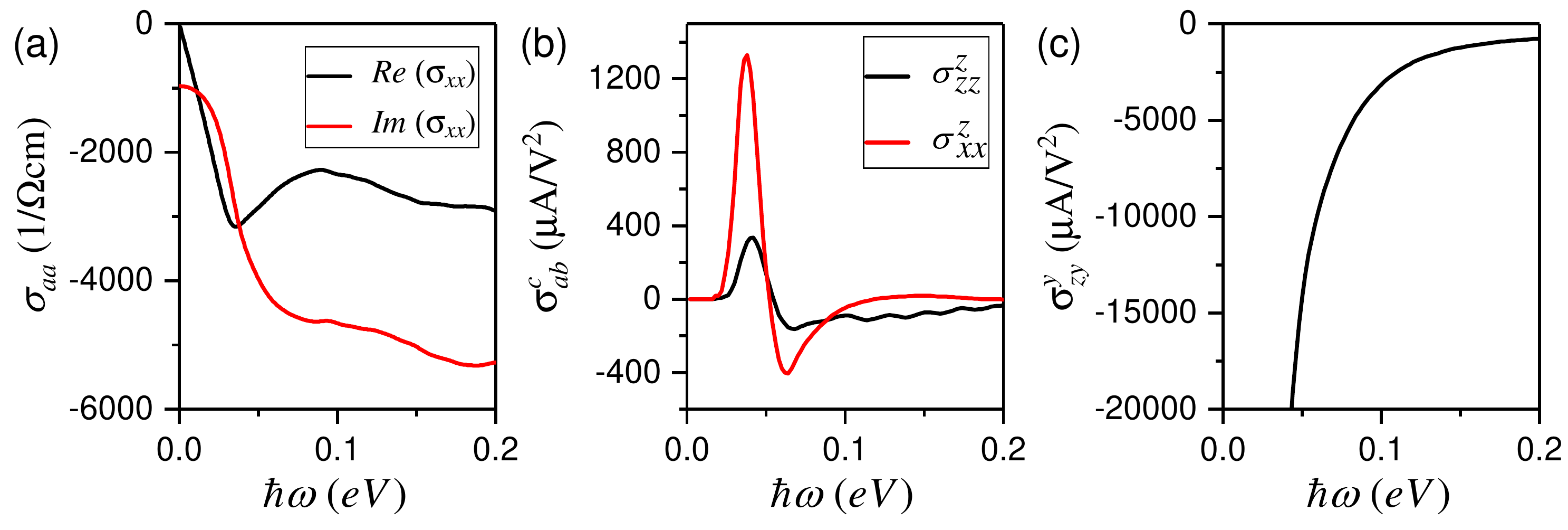}
\end{center}
\caption{
Photon energy dependence of
  (a) optical conductivity, (b) shift current conductivity under linearly
  polarized light, (c) circular
  photogalvanic conductivity.
}
\label{photon}
\end{figure}

Since the photocurrent response arises from both real and virtual band
transitions, it generally has a strong dependence on the incident photon energy. As we are starting from
a relaxation time approximation, the incident photon energy in our
calculation should be above $5$ meV
(the typical relaxation time $\delta=\frac{\hbar}{ \tau} $ for metallic system, we
use $\delta=10$ meV in our calculations).
Thus, we focus on the mid-infrared region from 20 meV to 200 meV,
which contains the transitions between Weyl bands.
In TaAs, two groups of type-I Weyl nodes exist
: (1) four pairs of WPs,
noted as $W_1$, on the $k_z=0$
plane with energy -23 meV; (2) eight pairs of WPs, noted as $W_2$, out
of $k_z =0$ plane with energy 14
meV. The shift current shown in Fig~\ref{photon}
has a strong peak at photon energy $\hbar \omega=40$
meV, around twice of the energy of in-plane Weyl nodes, and is almost zero below
the Weyl node energy scale in our calculation.
 This is explained by real transitions from band $n$ to band $l$ in
Fig.\ref{optical}, photocurrent is nonzero only when $E_{l}(\vec{k}) > 0,
E_n(\vec{k})<0, E_l(\vec{k})-E_n(\vec{k})=\hbar \omega$, and increase when $\hbar \omega$
is decreasing because of the $\frac{1}{\omega^2}$ in the prefactor of
Eq.\ref{second-kubo}.

For the photon energy dependence of CPGE, the $1/(\hbar \omega)^2$ behaviour is
observed in the region where our approach is valid. Since the energy denominator
$Re(D_1D_2$ is the dispersive part of the second order optical response, the
complex integral is nearly unchanged in low-frequency regime, leading to a
$1/(\hbar \omega)^2$ dependence due to the prefactor of Eq. \ref{second-kubo}

\textit{Effect of disorders and fluctuations.}
Next, we discuss the effect of temperature and impurity scattering to the
photocurrent generation.
In our calculation, the effect of disorder and fluctuations are taken into
account by the constant relaxation time $\tau$, which is not considered in the shift
vector formalism ~\cite{young2012first,Sipe2000}.  Since the distribution of
shift current in momentum space is quite concentrated around Weyl nodes, the
constant relaxation time would lead to almost the same results compared with
more realistic momentum dependent relaxation time.
Another possible effect on the conductivities comes from the change in the electron distribution. However, since most
of the experiments are carried out at low temperature ($k_BT=4.3$ meV ($T=50$K),
which is comparable to $\delta$ and much smaller than the frequency of light),
we expect the temperature change in the Fermi-Dirac distribution function does not modify the conductivity significantly.

\begin{figure}[htbp]
\begin{center}
\includegraphics[width=0.4\textwidth]{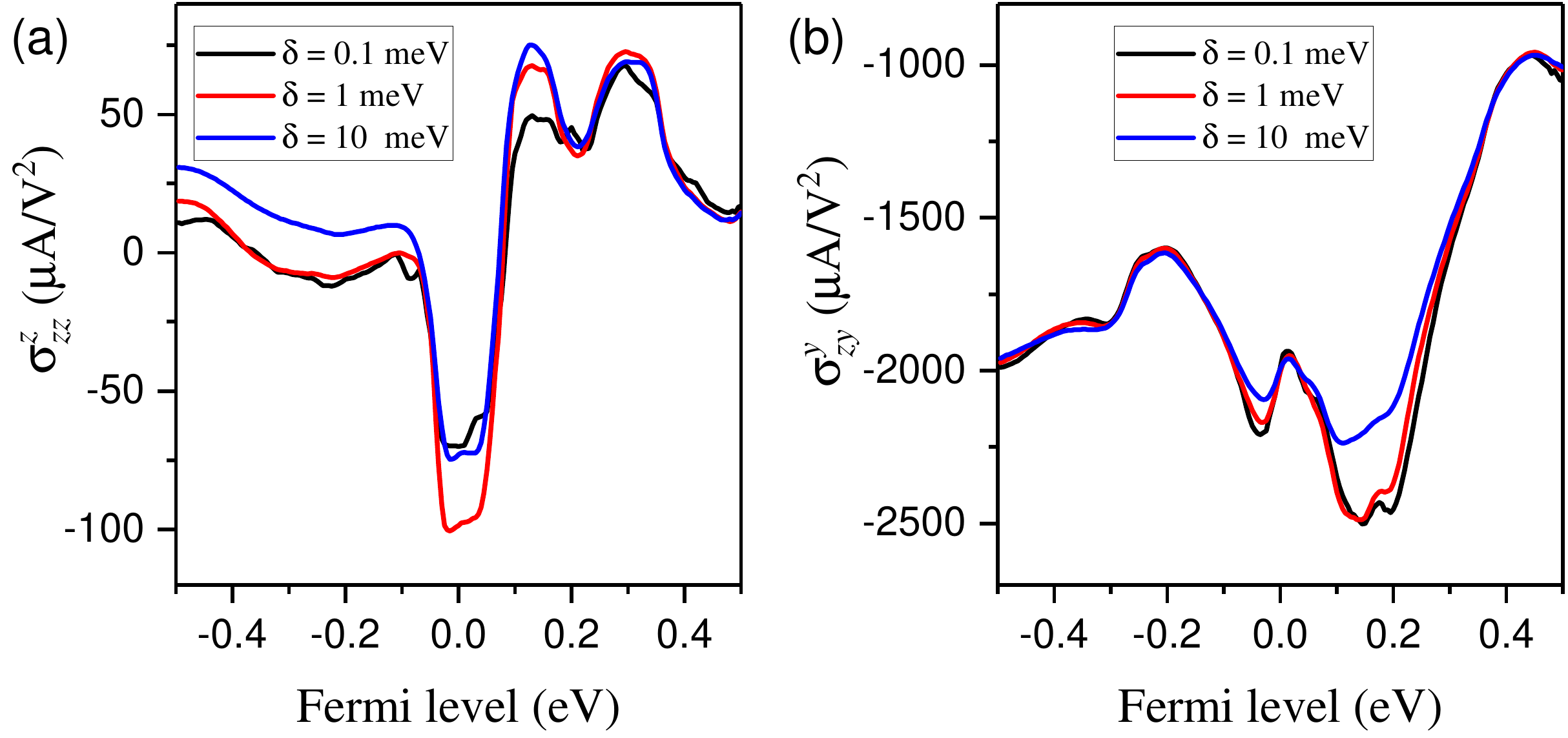}
\end{center}
\caption{
  Fermi level dependence of (a) shift current and (b) circular photogalvanic
  conductivity at $\hbar \omega=120$ meV. Each lines shows the results for different relaxation time.
}
\label{relaxation}
\end{figure}

Figure~\ref{relaxation} show the chemical potential dependence of shift current and CPGE, calculated with different relaxation time.
As shown in Fig.~\ref{relaxation}, both terms show
only small dependence to the relaxation time. For shift current
$\sigma^z_{zz}$, the response current is maximized when Fermi level is adjusted
around the Weyl nodes energy, and change only by $20\%$ even if the relaxation time is changed by a factor of 100.
For the CPGE $\sigma^z_{yz}$ curve, the response current is almost unchanged at the
charge neutrality point, and does not show strong dependence on the Weyl nodes
energy level.

\textit{Two- and three-band processes.}
  For given valence band $n$ and conduction band
  $l$, the CPGE and shift current should sum over the real transition ($n \rightarrow l$, $l=m$ in Eq. \ref{second-kubo})
  and also the virtual transitions ($n \rightarrow m \rightarrow l$, $l \neq m$ in Eq. \ref{second-kubo}) for all third bands $m$.
  To understand the importance
  of virtual transitions, here we separate the two- and three-band process contributions for
  the response at incident photon energy $\hbar \omega=120$ meV, to
  investigate which one is more essential in the photocurrent generation.

\begin{figure}[htbp]
\begin{center}
\includegraphics[width=0.4\textwidth]{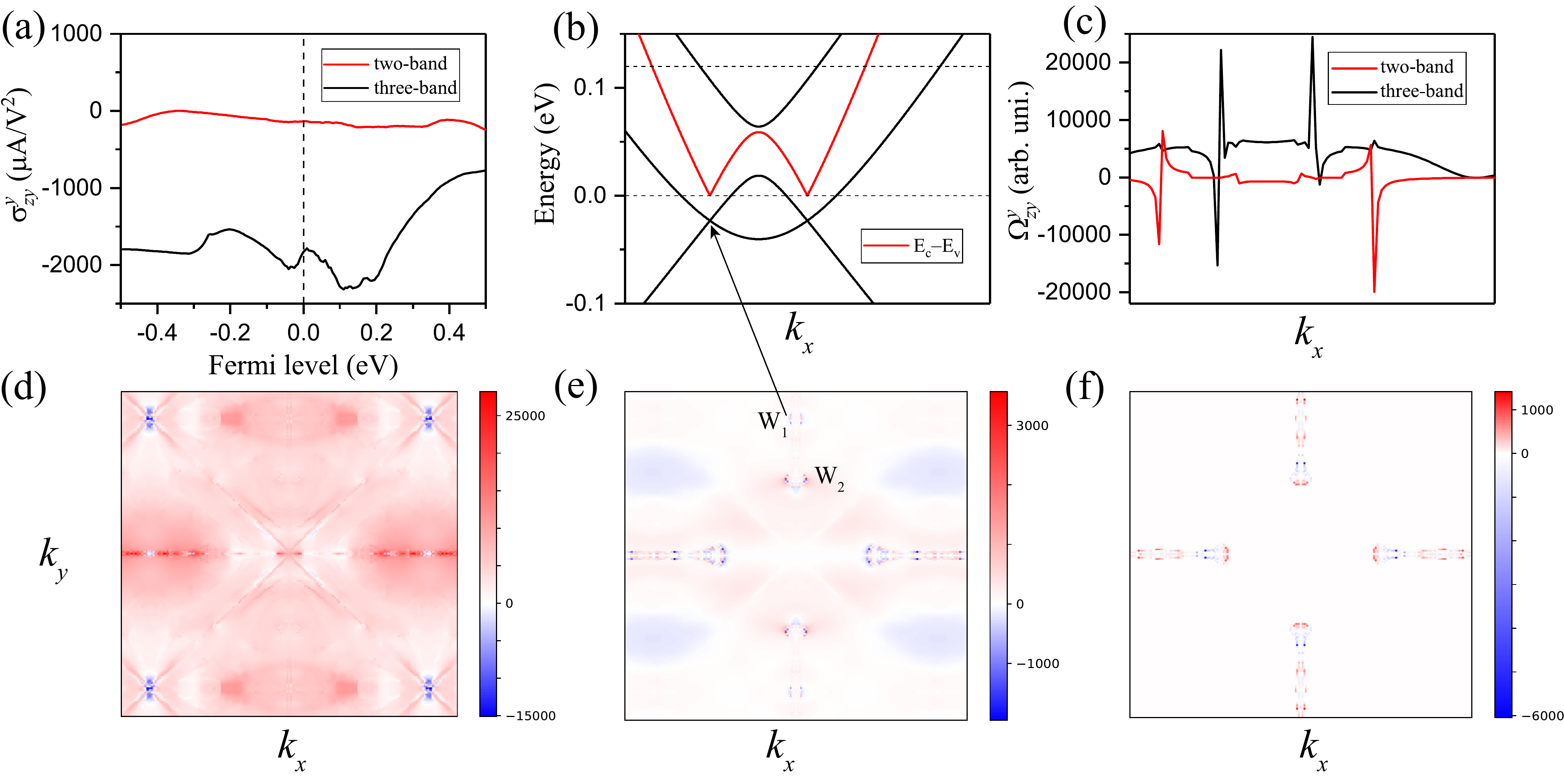}
\end{center}
\caption{
  (a) Three-band and two-band part of the Fermi level dependent CPGE
  $\sigma^y_{zy}$ at $\hbar \omega=120$ meV,
  (b) Band srtucture in $k_x$ direction across the $w_1$ Weyl node;
  (c) Three-band and two-band contribution of $\Omega^y_{zy}$
  (CPGE $\sigma^y_{zy}$ in momentum space)
  throught the same $k$ path as (b), blue curve is the band gap between valence
  and conduction band;
  (d) Three-band part of $\sigma^y_{zy}$(CPGE) in first Brillouin zone;
  (e) Two-band part of $\sigma^y_{zy}$(CPGE) in first Brillouin zone;
  (f) $\sigma^z_{zz}$(shift) in first Brillouin zone.
}
\label{inter-intra}
\end{figure}

  As shown in Fig.~\ref{inter-intra}(a), the three-band part of CPGE
  $\sigma^y_{zy}$ is $1825 \mu A/V^2$, while the two-band part is only $75 \mu
  A/V^2$ at the charge neutrality point.
  We obtain the $J_y = 1.2 \times 10^{-4}$ A with only two-band transitions in our method, which matches well
  with the theoretical calculated result $1.015 \times
  10^{-4}$ A in Ref.\cite{ma2017direct} via an effective two-band model.

  Similarly, for the entire range of Fermi level we calculated, a large contribution to the photocurrent comes from the three-band processes. The distribution of three-band contribtion for $\sigma^y_{zy}$
  is quite dispersed  in momentum space, in contrast to the of the two-band part concentrating around WPs.
  In total magnitude, the two-band process is ten times smaller than the three-band process.
  Taking a closer look into the small area around $W_1$ WPs, the two-band part solely comes from $E_v(\vec{k})-E_c(\vec{k})=120$ meV,
  which is the direct transition between two Weyl bands; while the three-band contribution stay
  almost uniformly in the momentum space, implying that virtual transitions have a larger contribution than the real transitions.

  It should be stressed that the shift current $\sigma^z_{zz}$ is purely a
  three-band process, as we have analyzed according to Eq. 1 and have also confirmed in numerical calculations.
    Therefore, it is necessary to include a third band for the evaluation of the photocurrent
  $\vec{J}$ parallel to electric field $\vec{E}$. In the momentum space
  distribution of shift current $\sigma^z_{zz}$, the nonzero part is
  concentrated around the WPs, which shows the absorptive nature of shift
  current.   Thus, we can conclude that shift current in Weyl system
  comes from the interplay of Weyl nodes and third trivial bands when the
  incident photon energy is at the same scale of the energy of Weyl nodes.

\textit{Discussion.}
We have systematically studied the photocurrent response both for
linearly and circularly polarized lights in type-I WSM TaAs, and show that shift
current spectrum has a strong dependence with Weyl points energy, while CPGE
shows a $1/(\hbar \omega)^2$ behaviour in mid infrared regime, when the incident
photon energy is larger than the smearing energy. Comparing
our calculated results with a recent photocurrent experiments, we observe
that the CPGE experiment of TaAs~\cite{ma2017direct} measured
$\sigma^y_{zy}(CPGE)$ 
with the incident photon energy $\hbar \omega=120$ meV.
Our calculated $\sigma^y_{zy}$ is $1900 ~\mu A/V^2$, gives a photocurrent
$J_y=2.1 \times 10^{-3} A$ under the experimental laser power.
Taking into account a scaling factor 10$^{-4}$ determined in experiment~\cite{ma2017direct}
and other unspecified decay channels, our results agrees well with the experimental value of $40 \times 10^{-9}$ A.
The calculated shift current $J_y$ is $8 \times
10^{-5}$ A in this setup ($4\%$ of the photocurrent from circularly polarized
light), which may possibly explain why shift current was neglected in Ref
.~\onlinecite{ma2017direct} that focused on the CPGE.

Recently the shift current was experimentally studied in TaAs~\cite{osterhoudt2017colossal}
and $\sigma^z_{xx}$,
$\sigma^z_{zz}$, and $\sigma^x_{zx}$ were measured at photon energy $\hbar \omega=117$
meV, which is at least an order of magnitude larger than previously measured
materials (e.g. $\sigma^z_{zz}(shift)=0.013 \mu A/V^2$ in
BaTiO$_3$ with visible light~\cite{zenkevich2014giant,young2012first}).
Our calculated $\sigma^x_{zx}(shift)$ is $79 \mu A/V^2$, in good agreement with
the experimental result $\sigma^x_{zx}= 26 \mu A/V^2$.

Apart from the above fixed photon energy experiments, it would be interesting
to investigate the frequency dependent photocurrent both for circularly and
linearly polarized light, to verify the $1/(\hbar \omega)^2$ dependence of CPGE and
the peak of shift current for $\hbar \omega$ being around twice of the WP ($W_1$) energy.

 In addition, the calculated SH susceptibility $\chi^z_{zz}$ and
 the ratio of
  $\chi^x_{zx}/\chi^z_{zz}$ are 6200 pm/V and 0.3 respectively,
  which are quite closed to the measured value 7200 pm/V and 0.031
  at low temperature~\cite{wu2017giant,subm}.

In summary, we have developed a first-principles multiband approach to determine the
photocurrent response from linearly and circularly polarized lights. We have established
that the virtual transitions from Weyl bands to trivial bands play an essential
role in the photocurrent generation process.
In general, our method is also useful to study the nonlinear optical responses
in ordinary metallic and insulating materials.


\bibliography{ref}
\end{document}